\documentclass{ws-ijbc}

\usepackage{graphicx}
\graphicspath{{pdf_figures/}}
\usepackage{multirow,bigstrut}
\usepackage{amssymb,amsmath,bm}

\newlength\figwidth
\usepackage{lineno}
\usepackage{color}
\definecolor{dgreen}{rgb}{0,.6,0}

\newlength\singleimagewidth
\newlength\twoimagewidth
\newlength\imagewidth
\begin{document}

\catchline{}{}{}{}{} 

\markboth{C. Li et al.}{Breaking an image encryption algorithm}

\title{Deciphering an image cipher based on mixed transformed Logistic maps}

\author{Yuansheng Liu}
\address{College of Information Engineering,\\
 Xiangtan University, Xiangtan 411105, Hunan, China}

\author{Hua Fan}
\address{Information Security Certification Center, Beijing 100020, China}

\author{Eric Yong Xie}
\address{College of Information Engineering,\\
Xiangtan University, Xiangtan 411105, Hunan, China}

\author{Ge Cheng }
\address{School of Mathematics and Computational Science,\\
Xiangtan University, Xiangtan 411105, Hunan, China}

\author{Chengqing Li}
\address{College of Information Engineering,\\
 Xiangtan University, Xiangtan 411105, Hunan, China\\
 DrChengqingLi@gmail.com}

\maketitle

\begin{history}
July 13, 2015
\end{history}

\begin{abstract}
Since John von Neumann suggested utilizing Logistic map as a random number generator in 1947,
a great number of encryption schemes based on Logistic map and/or its variants have been proposed.
This paper re-evaluates the security of an image cipher based on transformed logistic maps and proves that
the image cipher can be deciphered efficiently under two different conditions:
1) two pairs of known plain-images and the corresponding cipher-images with computational complexity of $O(2^{18}+L)$;
2) two pairs of chosen plain-images and the corresponding cipher-images with computational complexity of $O(L)$, where $L$ is the number of pixels in the plain-image. In contrast, the required condition in the previous deciphering method is eighty-seven pairs of chosen plain-images and the corresponding cipher-images with computational complexity of $O(2^{7}+L)$. In addition, three other security flaws existing in most Logistic-map-based ciphers are also reported.
\end{abstract}


\section{Introduction}

In the cyber era facing 5G (5th generation) mobile networks, all kinds of security problems about image
data are encountering serious challenges \cite{Li:Optimal:SP2011,Jian:Move:TIFS15}. The seeming similarity between chaos and cryptography promoted
their combination to design efficient and secure encryption schemes,
where one or more chaotic systems were adopted to determine position permutation
relationship \cite{Chen:Symmetric:CSF2004,Fridrich:Symmetric:IJBC1998}, generate pseudo-random number sequence (PRNS) \cite{Mao:Novel:IJBC2004,Zhu:Novel:OC2012,YuSM:feedback:IJBC2014,YCZhou:Chaotic:TC2015}, produce cipher-text directly \cite{Baptista:Lotistic:PLA98}
and construct public key encryption scheme \cite{Bose:public:PRL05}.
As an integral part of cryptology, security analysis on a given encryption scheme checks its real capability on achieving
balance point between security and the cost computational, and also provides another
perspective on studying property of the underlying theory \cite{KNUTH:congruential:IEEETIT1985,Alvarez:Some:IJBC2006}.
Some cryptanalytic works have shown that many chaos-based
encryption schemes have security problems of different extents from
modern cryptographical point of view \cite{Alvarez:Cryptanalysis:PLA2004,
Chen:Chosen:CSII2006, Solak:Cryptanalysis:IJBC2010, Li:Optimal:SP2011,
Li:hyperchaotic:ND2013,Cqli:breakmodulo:IJBC13}.

Logistic map is one of the most famous chaotic systems. It comes from
discrete quadratic recurrence form of the Logistic equation,
a model of population growth first published by P. Verhulst in 1845.
The application of Logistic map in cryptology
can be traced back to John von Neumann's suggestion on utilizing it as a random number generator in 1947 \cite{Neumann:logistic:BAMS47}.
The map becomes very popular after the biologist Robert May used it as a discrete-time demographic model in 1976 \cite{MAY:Logistic:Nature1976}.
Due to simple form and relatively complex dynamical properties of Logistic map, it was extensively used to design encryption
schemes or generate PRNS \cite{Baptista:Lotistic:PLA98,Phatak:LogisticRNG:PRE95,Kocarev:Logistic:PLA01,Jakimoski:Chaos:CSI2001}. Even in \textit{Web of Science}, one can find that more than two hundred papers on application of Logistic map in cryptography were published between 1998 and 2014.
Among them, a few papers reported some security deficiencies specially caused by Logistic map, such as estimation of control parameter from neighboring states \cite{Li:AttackingRCES2008}, short period of the states orbit \cite{Persohn:AnalyzeLogistic:CSF12,Li:LogisticPRNG:ITVLSIS2012},
stable distribution of estimation error of the control parameter \cite{Li:logistic:ND2014}. To tackle the flaws, various remedies were proposed, such as modify Logistic map itself \cite{Sam:TLM:MTA2012} or postprocess the raw chaotic states \cite{Li:LogisticPRNG:ITVLSIS2012}.

In \cite{Sam:TLM:MTA2012}, a novel image cipher based on
mixed transformed Logistic maps (MTLM) was proposed, where
the modulo addition and the XOR operations
are employed in diffusion procedure, which are all controlled by
PRNS generated by iterating MTLM. Essentially, the image cipher falls in the categories of encryption schemes based on function
\begin{equation}
\label{eq:essentialfunction}
y = (\alpha \dotplus x)\oplus (\beta \dotplus x),
\end{equation}
where $y$, $\alpha$, $\beta$, and $x$ are $n$-bit non-negative integers,
$\alpha \dotplus x = (\alpha + x) \mod 2^{n}$, and $\oplus$ denotes
the eXclusive OR (XOR) operation. Detailed cryptographic properties of Eq.~(\ref{eq:essentialfunction}) have been given in \cite{Cqli:breakmodulo:IJBC13}.
Recently, YS Zhang et al. found that the cipher is insecure against
chosen-plaintext attack and the equivalent secret key
can be obtained by eighty-seven pairs of
chosen plain-images \cite{Zhang:Cryptanalyzing:MTA2014}.

This paper re-evaluates the security of the image cipher proposed
in \cite{Sam:TLM:MTA2012}, and points out the following main insecurity issues:
1) the deciphering performance
of the chosen-plaintext attack can be further
improved, in terms of both the number of required plaintexts and computationalal complexity;
2) the cipher can be broken efficiently with only two pairs of
known plain-images and their corresponding cipher-images;  3) the image cipher suffers other security flaws like insensitivity with respect to change of plain-image/secret key and weak randomness of the used PRNS.

The remaining of the paper is organized as follows. The next section gives a brief introduction of the image
cipher under study. Then, the comprehensive cryptanalyses on it are presented in Sec.~\ref{sec:ca},
together with some experimental results. Finally, the last section concludes the paper.

\section{Description of the image cipher under study}
\label{sec:alogrithm}

The plain-image of the image cipher under study is a RGB color image of
size $H\times W$ (height $\times$ width), which can be
represented as an 8-bit integer matrix of size $3\times L$,
$\mathbf{I}=\left\{I(i)\right\}_{i=1}^{L}=\left\{(R(i), G(i), B(i)\right)\}_{i=1}^{L}$, by
scanning the pixels in the raster order, where $L=H\cdot W$.
Similarly, the corresponding cipher-image is denoted by
$\mathbf{I}'=\{I'(i)\}_{i=1}^{L}=\{(R'(i), G'(i), B'(i))\}_{i=1}^{L}$.
Then, the four main parts of the image cipher under study are described as follows\footnote{To simplify the description of the image cipher under study,
some notations in the original paper \cite{Sam:TLM:MTA2012}
are modified under the condition that the essential form kept unchanged.}.
\begin{itemize}
\item
\textit{The secret key} is composed of six odd integers
$\{r_u\}_{u=1}^{6}$ and three control
parameters $k_1$, $k_2$, $k_3$, initial state $(x_0, y_0, z_0)$
of MTLM proposed in
\cite{Sam:TLM:MTA2012}, which is given as
\begin{equation*}
\begin{cases}
x_{i+1}=(3.735\cdot k_1 \cdot (1+x_i)^2 \cdot \sin(1/(1+y_i^2)) )  \bmod{1},         \\
y_{i+1}=(3.536\cdot k_2 \cdot x_{i+1}   \cdot \sin(x_{i+1}\cdot  y_i)  (1+z_i^2) )\bmod{1},    \\
z_{i+1}=(3.838\cdot k_3 \cdot x_{i+1}   \cdot (1 + y_{i+1}\cdot z_i) ) \bmod{1},
\end{cases}
\end{equation*}
where $r_u \in [0, 256]$, $\left|k_1\right|>37.7$, $\left|k_2\right|>39.7$ and $\left|k_3\right| > 37.2$.

\item
\textit{Keystream generation procedure}:
Iterate the above MTLM $L$ times to obtain a chaotic states sequence
$\{(x_i, y_i, z_i)\}_{i=1}^{L}$. Then, generate keystream as follows: for $i=1 \sim L$, set
\begin{equation}
\nonumber
\begin{cases}
X_{i} = \lfloor 256 \cdot x_{i} \rfloor, \\
Y_{i} = \lfloor 256 \cdot y_{i} \rfloor, \\
Z_{i} = \lfloor 256 \cdot z_{i} \rfloor,
\end{cases}
\end{equation}
where $\lfloor x \rfloor$ quantizes $x$ to the nearest integer less than or
equal to $x$.

\item\textit{The encryption procedure} consists of
the following three operations.
\begin{itemize}
\item
\textit{Initial permutation}: For $i=1\sim H, j=1\sim W$, set
\begin{equation}
\label{eq:permutation}
\begin{cases}
R^{\dagger}((i-1)\cdot W+ j) = R((t_1-1)\cdot W+ t_2), \\
G^{\dagger}((i-1)\cdot W+ j) = G((t_3-1)\cdot W+ t_4), \\
B^{\dagger}((i-1)\cdot W+ j) = B((t_5-1)\cdot W+ t_6),
\end{cases}
\end{equation}
where
\begin{equation}
\nonumber
t_u =
\begin{cases}
1 + (31 \cdot i\cdot r_u)\bmod{H}, &\mbox{if } u\in\{1, 3, 5\}; \\
1 + (31 \cdot j\cdot r_u)\bmod{W}, &\mbox{otherwise},
\end{cases}
\end{equation}
i.e., $t_u$ is a function of $i$ or $j$ according to the value of $u$.

\item
\textit{Nonlinear diffusion}: For $i = 1 \sim L$, set
\begin{equation}
\label{eq:nonlinear}
\begin{cases}
R^{\ddagger}(i) = \left(\left(R^{\dagger}(i)\ggg 4\right) \dotplus X_{i}\right)\oplus Y_{i}, \\
G^{\ddagger}(i) = \left(\left(G^{\dagger}(i)\ggg 4\right) \dotplus X_{i}\right)\oplus Y_{i}, \\
B^{\ddagger}(i) = \left(\left(B^{\dagger}(i)\ggg 4\right) \dotplus X_{i}\right)\oplus Y_{i}, \\
\end{cases}
\end{equation}
where $a\ggg 4 = 16\cdot(a\bmod{16})+\lfloor a/16 \rfloor$.

\item \textit{Zigzag diffusion}: 1) re-scan all pixels of $\mathbf{I}^{\ddagger}=\{I^{\ddagger}(i)\}_{i=1}^{L}=\left\{(R^{\ddagger}(i), G^{\ddagger}(i), B^{\ddagger}(i))\right\}_{i=1}^{L}$ in the zigzag order and still store the result with $\mathbf{I}^{\ddagger}$;
    2) encrypt each element of $\mathbf{I}^{\ddagger}$ in order by
\begin{equation}
\label{eq:zigzag}
\begin{cases}
R'(i) = R^{\ddagger}(i) \oplus R'(i-1) \oplus Z_{i}, \\
G'(i) = G^{\ddagger}(i) \oplus G'(i-1) \oplus Z_{i}, \\
B'(i) = B^{\ddagger}(i) \oplus B'(i-1) \oplus Z_{i},
\end{cases}
\end{equation}
where $R'(0) = G'(0) = B'(0) = 0$.
\end{itemize}

\item \textit{Decryption procedure} is similar to the encryption one except the following points:
\begin{itemize}
\item the above encryption operations are run in a reverse order;
\item Eqs.~\eqref{eq:permutation}, \eqref{eq:nonlinear} and \eqref{eq:zigzag} are replaced by
\begin{equation}
\nonumber
\begin{cases}
R((t_1-1)\cdot W+ t_2) = R^{\dagger}((i-1)\cdot W+ j), \\
G((t_3-1)\cdot W+ t_4) = G^{\dagger}((i-1)\cdot W+ j), \\
B((t_5-1)\cdot W+ t_6) = B^{\dagger}((i-1)\cdot W+ j),
\end{cases}
\end{equation}
\begin{equation}
\label{eq:nonlinear_de}
\begin{cases}
R^{\dagger}(i) = \left(\left(\left(R^{\ddagger}(i)\oplus Y_{i}\right) - X_{i}\right)\bmod{256}\right)\ggg 4, \\
G^{\dagger}(i) = \left(\left(\left(G^{\ddagger}(i)\oplus Y_{i}\right) - X_{i}\right)\bmod{256}\right)\ggg 4, \\
B^{\dagger}(i) = \left(\left(\left(B^{\ddagger}(i)\oplus Y_{i}\right) - X_{i}\right)\bmod{256}\right)\ggg 4, \\
\end{cases}
\end{equation}
and
\begin{equation}
\label{eq:zigzag_de}
\begin{cases}
R^{\ddagger}(i) = R'(i)\oplus R'(i-1) \oplus Z_{i}, \\
G^{\ddagger}(i) = G'(i)\oplus G'(i-1) \oplus Z_{i}, \\
B^{\ddagger}(i) = B'(i)\oplus B'(i-1) \oplus Z_{i},
\end{cases}
\end{equation}
respectively.
\end{itemize}
\end{itemize}
Observing Eqs.~(\ref{eq:nonlinear_de}), (\ref{eq:zigzag_de}), one can see
that sequences $\{X_{i}\}_{i=1}^{L}$ and $\{Y_{i}\oplus Z_{i}\}_{i=1}^{L}$ are the equivalent key of the two cascaded parts, \textit{Nonlinear diffusion} and
\textit{Zigzag diffusion}.

\section{Cryptanalysis}
\label{sec:ca}

To make cryptanalysis on the image cipher under study more complete, we first briefly
review previous deciphering method proposed by Zhang et al. Then, a best improvement of the attack
is presented in Sec.~\ref{ssec:ocpa}. Furthermore, we proposed an effective known-plaintext attack in Sec.~\ref{ssec:kpa}.
Finally, some security flaws of the scheme are presented.

\subsection{Chosen-plaintext attack proposed by Zhang et al.}
\label{subsec:cpa:zhang}

Chosen-plaintext attack is an attack model assuming that the attacker owns right to modify
plaintext and observes the corresponding ciphertext. Assume a plain-image $\mathbf{I}_1=\left\{I_1(i)\right\}_{i=1}^{L}=\left\{(R_1(i), G_1(i), B(i)\right)\}_{i=1}^{L}$ and its corresponding cipher-image $\mathbf{I}_1'=\{I'_1(i)\}_{i=1}^{L}=\{(R_1'(i)$, $G_1'(i), B_1'(i))\}_{i=1}^{L}$ are available.
Obviously, any permutation operation become invalid with respect to permutation object of a fixed value. To decipher the three encryption operations with the strategy of
\textit{Divide and Conquer}, the plain-image is chosen as $\mathbf{I}_1=\{(R_1(i)\equiv 0,$ $G_1(i)\equiv d, B(i))\}_{i=1}^{L}$, where $d\in \{0, \ldots, 255\}$. According to Eqs.~\eqref{eq:nonlinear} and \eqref{eq:zigzag}, one has
\begin{equation}
\label{eq:value0}
R_1'(i) \oplus R_1'(i-1) = X_{i} \oplus Y_{i} \oplus Z_{i},
\end{equation}
and
\begin{equation}
\label{eq:valued}
G_1'(i) \oplus G_1'(i-1) = \left(\left(d \ggg 4\right) \dotplus X_{i}\right) \oplus Y_{i} \oplus Z_{i}.
\end{equation}
Incorporate $(Y_{i} \oplus Z_{i})$ in Eq.~(\ref{eq:value0}) into Eq.~\eqref{eq:valued}, one further has
\begin{linenomath}
\begin{equation}
\label{eq:zhangcore}
R_1'(i) \oplus R_1'(i-1) \oplus G_1'(i) \oplus G_1'(i-1)
= \left(\left(d \ggg 4\right) \dotplus X_{i}\right) \oplus X_{i}.
\end{equation}
\end{linenomath}
The above equation can be attributed to a special case of Eq.~(\ref{eq:essentialfunction}),
\begin{linenomath}
\begin{equation}
\nonumber
y = (\alpha \dotplus x) \oplus x.
\end{equation}
\end{linenomath}
As shown in Table~2 of \cite{Cqli:breakmodulo:IJBC13}, only the scope of possible values of $X_{i}$
can be obtained by solving one set of Eq.~(\ref{eq:zhangcore}). Given more pairs of chosen plain-images
and the corresponding cipher-images, the scope would become more and more narrow, and all the seven least significant
bits of $X_i$ may be confirmed. In \cite{Zhang:Cryptanalyzing:MTA2014}, Zhang et al. claimed that $256$ plain-images of different fixed values can reveal
$\{X_{i}\}_{i=1}^{L}$ (As the relation shown in Eq.~(\ref{eq:cpa:equi}), the most significant bit of $X_{i}$ is ignored). As a RGB color image has three channel components,
one can conclude that the chaotic keystream $\{X_{i}\}_{i=1}^{L}$ can
be reconstructed with $\lceil 256/3\rceil = 86$ pairs of chosen
plain-images. Then, the keystream
$\{Y_{i}\oplus Z_{i}\}_{i=1}^{L}$ can be obtained by Eq.~\eqref{eq:value0}.
Then, only the \textit{Initial permutation} is left. One can use one more chosen plain-image, e.g., modified version of any previous chosen-image by
changing one single pixel value, to recover the parameters of the position permutation part, $\{r_u\}_{u=1}^{6}$,
via Eq.~\eqref{eq:permutation}.

In all, the essential idea of Zhang et al.'s attack is to narrow the scope of
$x$ by verifying Eq.~(\ref{eq:essentialfunction}) of $\beta\equiv 0$ with different sets of $(y, \alpha)$.
As $y \in [0, 255]$, they enumerate all possible values
of $\alpha \in [0, 127]$, and then recover $x$ by observing the distribution of $y$.
So, the required number of chosen plain-image is eighty-seven
and the computational complexity is $O(128 \cdot L)$.

\subsection{Optimum chosen-plaintext attack}
\label{ssec:ocpa}

Based on the above discussion, this subsection presents an improved chosen-plaintext attack
based on the following proposition.
\begin{proposition}
Assume that $\alpha, \beta$, and $x$ are all $n$-bit integers, then a lower bound on the number of queries $(\alpha, \beta)$ to solve
Eq.~(\ref{eq:essentialfunction}) in terms of modulo $2^{n-1}$ for any $x$ is 1 if $n=2$; 2 if $n>2$.
\end{proposition}

Following the above Proposition, a corollary listed two typical sets of $(\alpha, \beta)$
to determine $x$ in Eq.~\eqref{eq:essentialfunction} when $n=8$, where
the two queries are $(0,170)$ and $(170,85)$. Thus, one can choose a plain-image
$\mathbf{I} =\{(R(i)\equiv 0$,
$G(i)\equiv 170, B(i))\equiv 85\}_{i=1}^{L}$, and then one further has
\begin{linenomath}
\begin{equation}
\nonumber
\begin{cases}
R'(i) \oplus R'(i-1) \oplus G'(i) \oplus G'(i-1) =  \\
\hspace{4cm}\left(0 \dotplus X_{i}\right)
 \oplus \left(170 \dotplus X_{i}\right),\\
G'(i) \oplus G'(i-1) \oplus B'(i) \oplus B'(i-1) =  \\
\hspace{4cm} \left(170 \dotplus X_{i}\right)
\oplus \left(85 \dotplus X_{i}\right),
\end{cases}
\end{equation}
\end{linenomath}
from Eqs.~\eqref{eq:nonlinear} and \eqref{eq:zigzag}.
Thus, $\{X_{i}\}_{i=1}^{L}$ can be revealed with the above chosen plain-image.
Once $\{X_{i}\}_{i=1}^{L}$ is recovered,
only one more chosen plain-image is required to break \textit{ Initial permutation} as above. Therefore, the equivalent key of
the image cipher under study can be revealed with only two chosen plain-images and the computational complexity of
the improved chosen-plaintext attack is only $O(L)$. As shown in \cite{Cqli:breakmodulo:IJBC13}, it is impossible to
solve $x$ of most values in Eq.~(\ref{eq:essentialfunction}) with only one set of $(\alpha, \beta)$. So, this attack method can be considered
as optimum.

\subsection{Known-plaintext attack}
\label{ssec:kpa}

The known-plaintext attack is a weaker version of the chosen-plaintext attack
as the attacker can not modify the plaintext. So, the former is more important for security analysis.

\begin{figure}[!htb]
\centering
\includegraphics[width=\singleimagewidth]{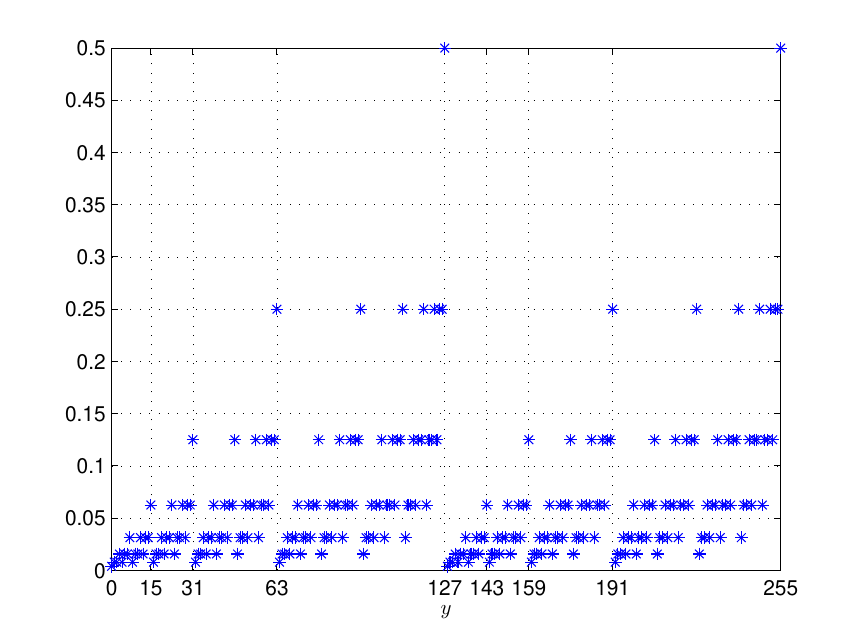}
\caption{The probability of $(\alpha, \beta)$ passing verification
of Eq.~\eqref{eq:essentialfunction} under different values of $y$.}
\label{fig:prob}
\end{figure}

Assume another plain-image $\mathbf{I}_2 =  \{I_2(i)\}_{i=1}^{L}$ $=\{(R_2(i)$, $G_2(i), B_2(i))\}_{i=1}^{L}$,
and the corresponding cipher-image
$\mathbf{I}'_2=\left\{I'_2(i)\right\}_{i=1}^{L}=\left\{\left(R'_2(i), G'_2(i), B'_2(i)\right)\right\}_{i=1}^{L}$,
are available.
According to Eqs.~\eqref{eq:nonlinear} and \eqref{eq:zigzag},
one has
\begin{multline}
\label{eq:essentialfunction:extone}
R'_1(i) \oplus R'_1(i-1) \oplus R'_2(i) \oplus R'_2(i-1) =
 \left(\left(R^{\dagger}_1(i)\ggg 4\right) \dotplus X_{i}\right)
 \oplus \left(\left(R^{\dagger}_2(i)\ggg 4\right) \dotplus X_{i}\right).
\end{multline}
Referring to Proposition 2 in \cite{Li:hyperchaotic:ND2013}, one has
\begin{linenomath}
\begin{equation}
\label{eq:cpa:equi}
(a \dotplus (X_{i} \oplus 128)) = (a \dotplus X_{i}) \oplus 128.
\end{equation}
\end{linenomath}
Thus, one can see that $X_{i}$ is equivalent to $(X_{i}\oplus 128)$
in terms of existence of Eq.~\eqref{eq:essentialfunction:extone}.
According to Eq.~\eqref{eq:permutation}, one has
\begin{linenomath}
\begin{equation}
\nonumber
R^{\dagger}(i\cdot W) = R(u_i),
\end{equation}
\end{linenomath}
where $u_i = (((31 \cdot i\cdot r_1)\bmod{H})\cdot W+1)$ and $i \in \{1,\ldots, H\}$.
Referring to Eq.~\eqref{eq:essentialfunction:extone}, one further has
\begin{multline}
\label{eq:cpa:breakingper}
R'_1(i\cdot W) \oplus R'_1(i\cdot W-1) \oplus R'_2(i\cdot W) \oplus R'_2(i\cdot W-1)= \\
\left(\left(R_1(u_i)\ggg 4\right) \dotplus X_{(i\cdot W)}\right)\oplus \left(\left(R_2(u_i)\ggg 4\right) \dotplus X_{(i\cdot W)}\right).
\end{multline}
As $r_1$ is an odd integer and $r_1\in [0, 256]$,
there are only $128$ possible values. Thus,
one can enumerate the possible values of $r_1$ and then verify them by checking whether $X_{(i\cdot W)} \in \{0,\ldots, 127\}$
satisfies Eq.~\eqref{eq:cpa:breakingper} for any $i \in\{1, \cdots, H\}$.
If all the verifications pass, the remaining value of $r_1$ is considered as the right sub-key.
Apparently, the computational complexity of the search procedure is
$O(128\times H\times 128) = O(2^{14}H)$,
and the success of this method is determined by
the verification of Eq.~\eqref{eq:cpa:breakingper}.
The probability of $(\alpha, \beta)$ passing verification of
Eq.~\eqref{eq:essentialfunction} under different values of $y$ are shown
in Fig.~\ref{fig:prob}. Assume that the elements of $\left\{R_1(i\cdot W+1)\right\}_{i=1}^{H}$,
and $\left\{R_2(i\cdot W+1)\right\}_{i=1}^{H}$ follow uniform distribution,
one can assure that the probability of a wrong version of $r_1$ passing
the verification procedure is less than ${(1/2)}^{H}$, which means
the value of $r_1$ can be successfully recovered with only three times verification with an extremely high probability.
In the same way, we can obtain other five odd integers $\{r_i\}_{i=2}^{6}$.
Therefore, the computational complexity of deciphering the \textit{Initial permutation}
is $O(6 \times 128\times 3\times 128) = O(9\cdot 2^{15})\doteq O(2^{18})$.

Once the permutation part has been deciphered,
some bits of $\{X_{i}\}_{i=1}^{L}$ can be revealed with
even one pair of known plain-image and the corresponding cipher-image.
In the current situation, Eq.~(\ref{eq:zhangcore}) is replaced by
\begin{equation*}
\label{eq:essentialfunction:rg}
R'(i) \oplus R'(i-1) \oplus G'(i) \oplus G'(i-1) =
 \left(\left(R^{\dagger}(i)\ggg 4\right) \dotplus X_{i}\right)
 \oplus \left(\left(G^{\dagger}(i)\ggg 4\right) \dotplus X_{i}\right).
\end{equation*}
Obviously, the above equation falls in the general form of Eq.~\eqref{eq:essentialfunction}.
In \cite[Sec.~3.2]{Cqli:breakmodulo:IJBC13}, Li et al. proved that $\Pr(0) = 0.5$, $\Pr(1) = 0.4062$, $\Pr(2) = 0.3818$, and
$\Pr(i) \approx 0.37$ for $i > 3$, where $\Pr(i)$ denotes the probability that the $i$-th bit can be confirmed.
For each pair of known plain-image and the corresponding cipher-image, there are three equations of the form of Eq.~\eqref{eq:essentialfunction}.
When two known
plain-images are available, one can obtain ${6 \choose 2} = 15$ equations.
Therefore, one can assure that most bits of $\{X_{i}\}_{i=1}^{L}$ can be obtained with a high probability, which is larger
than $1-(1-0.37)^{15}=1-0.43^{15}$. It is easy to conclude that the computational complexity on confirming bits of $\{X_{i}\}_{i=1}^{L}$ is $O(L)$.

\begin{figure}[!htb]
\centering
\begin{minipage}[t]{\twoimagewidth}
\centering
\includegraphics[width=\twoimagewidth]{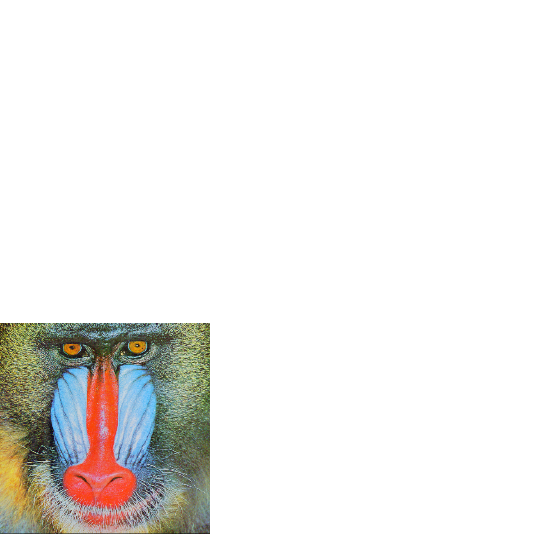}
a)
\end{minipage}
\begin{minipage}[t]{\twoimagewidth}
\centering
\includegraphics[width=\twoimagewidth]{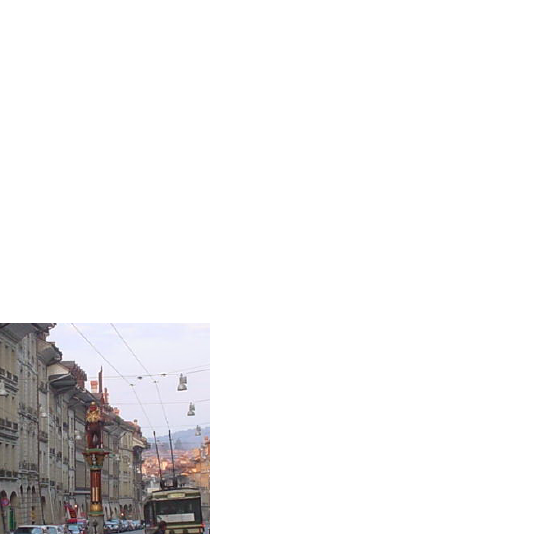}
b)
\end{minipage}
\\
\begin{minipage}[t]{\twoimagewidth}
\centering
\includegraphics[width=\twoimagewidth]{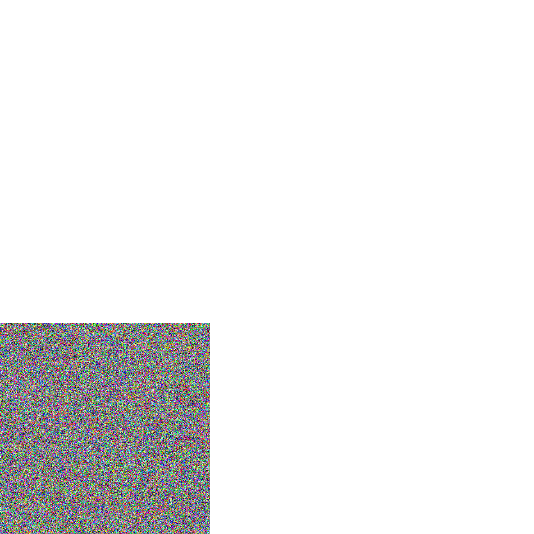}
c)
\end{minipage}
\begin{minipage}[t]{\twoimagewidth}
\centering
\includegraphics[width=\twoimagewidth]{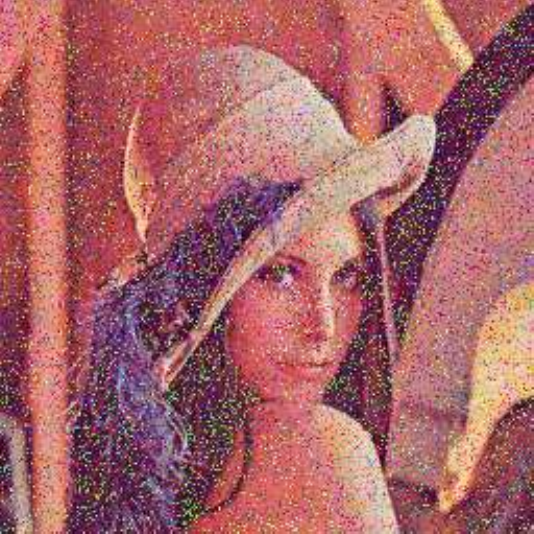}
d)
\end{minipage}
\begin{minipage}[t]{\twoimagewidth}
\centering
\includegraphics[width=\twoimagewidth]{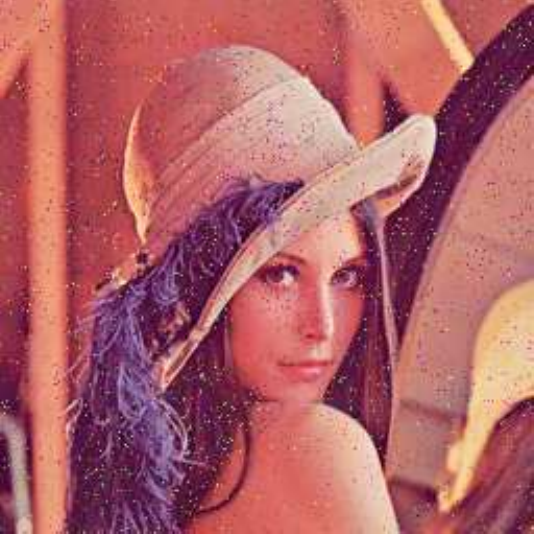}
d)
\end{minipage}
\caption{Known-plaintext attack:
a) the known plain-image ``Baboon";
b) the known plain-image ``Street";
c) the cipher-image of a plain-image ``Lenna";
d) the decryption result that the equivalent secret key of diffusion part
is reconstructed with the known plain-image ``Baboon";
e) the decryption result of c) with two known plain-images in a) and b).}
\label{figure:kp}
\end{figure}

To verify the real performance of the above known-plaintext attack,
a great number of experiments were performed with plain-images of size
$256\times 256$. Here, a typical example is shown,
where the secret key
$\{r_i\}_{i=1}^{6} = \{123, 57, 67, 89, 253, 221\}$,
$k_1 = 38.583, k_2 = 41.135, k_3 = 39.846$, and
$(x_0, y_0, z_0) = (0.485, 0.913, 0.751)$.
When only the plain-image ``Baboon" shown in Fig.~\ref{figure:kp}a)
is used to recover the approximate version of the equivalent secret key of the image cipher under study,
$(\{r_i\}_{i=1}^{6}$, $\{X_{i}\}_{i=1}^{L}, \{Y_{i}\oplus Z_{i}\}_{i=1}^{L})$, the decryption
result on another cipher-image shown in Fig.~\ref{figure:kp}c) is shown in Fig.~\ref{figure:kp}d).
When another plain-image shown in Fig.~\ref{figure:kp}b) is used together, the decryption
result on the cipher-image is shown in Fig.~\ref{figure:kp}e). It is counted that $26.53\%$ and $83.96\%$
of the pixels of the images shown in Fig.~\ref{figure:kp}d)
and Fig.~\ref{figure:kp}e) are correct.
As there is strong redundancy existing in neighboring pixels of image and human eyes owns strong robustness against noise in image \cite{Zhu:assessment:CASVT2015}, we can even observe some important visual information from Fig.~\ref{figure:kp}d) by naked eyes.
Therefore, one can conclude that two known plain-images can achieve a satisfactory deciphering performance.

\subsection{Other security flaws}

In this subsection, we list three other security flaws existing widely in Logistic-map-based ciphers, which
are all reduce the complexity of deciphering the ciphers seriously.
\begin{itemize}
\item \textit{Low sensitivity with respect to change of secret key:}

From the cryptographical point of view, a good secure image cipher should be sensitive to the secret key \cite{Schneier:Applied:2007}.
In \cite[Sec.~4.3]{Sam:TLM:MTA2012}, the author claimed that
the cipher under study has a great sensitivity to the secret key
based on some limited test results. Unfortunately, we found that
the image cipher under study fails to satisfy this security principle.
From the previous analysis, one can see that
$\{X_{i}\}_{i=1}^{L}$ and
$\{Y_{i}\oplus Z_{i}\}_{i=1}^{L}$ are the
equivalent key of Nonlinear and Zigzag diffusion parts. When
$Y'_{i}\oplus Z'_{i} = Y_{i}\oplus Z_{i}$,
then $Y'_{i}$ and $Z'_{i}$ are equivalent to $Y_{i}$ and
$Z_{i}$, respectively. Moreover, according to Eq.~\eqref{eq:cpa:equi},
$\left(X_{i}, Y_{i}\oplus Z_{i}\right)$ is equivalent
to $\left(X_{i}\oplus 128, Y_{i}\oplus Z_{i}\oplus 128\right)$
with respect to the encryption/decryption procedure, which means different keys
may successfully decrypt cipher-images encrypted with different secret keys.

\item \textit{Low sensitivity with respect to change of plaintext:}

Another cryptographical property required by a good
cipher (not visual cipher) is the avalanche effect, i.e., the ciphertexts
of two plaintexts with a slight change (e.g., only one pixel or bit is modified)
should be very dramatically different \cite{Schneier:Applied:2007}.
However, the image cipher under study is actually far away
from the property. From the encryption procedure,
there is only zigzag diffusion operation which can spread
the change to influence cipher-image,
and the change of one pixel of plain-image can only influence
pixels after the present pixel with the zigzag order.
For example, assume $R^{\dagger}(L-1)$ is permutated from $R(i')$.
If the value of $R(i')$ is modified, only three cipher pixels
$R'(L-W-1)$, $R'(L-1)$ and $R'(L)$ will be changed. This flaw is very important for
protecting image since a plain-image and its watermarked version may encrypted together.

\item \textit{Insufficient randomness of the keystream:}

In \cite[Sec.~2.2]{Sam:TLM:MTA2012}, it was claimed
that MTLM does not have security
issues existing in Logistic map. Li at al.
\cite{Li:AttackingBitshiftXOR2007} have shown that the
randomness of pseudo-random bit sequences derived from
the Logistic map is very weak. To further test the
randomness of the keystream generated by MTLM, we tested $100$ keystreams of
length $256\times 256\times 3 = 196608$ by using the NIST
statistical test suite \cite{Rukhin:TestPRNG:NIST10}.
The $100$ keystreams were generated with randomly selected
secret keys. For each test, the default significance level
$0.01$ was adopted. The results are shown in Table \ref{tab:random},
from which one can see that the keystream is not random enough.
\begin{table}[htbp]
\tbl{The performed tests with respect to a significance level $0.01$
and the number of sequences passing each test in $100$ randomly generated sequences.}
{\begin{tabular}{c|c}
\toprule
Name of Test & Number of Passed Sequences \\\hline
Approximate Entropy ($m = 10$) & $0$\\\hline
Block Frequency ($m = 128$) & $0$\\\hline
Cumulative Sums (Forward/Reverse) & $0/0$\\\hline
FFT & $100$\\\hline
Frequency & $0$\\\hline
Longest Run of Ones ($m = 10000$) & $99$\\\hline
Non-overlapping Template ($m = 9, B = 000000001$) & $0$\\\hline
Random Excursions ($x = 1$) & $0$\\\hline
Rank & $99$\\\hline
Runs & $0$\\\hline
Serial ($m = 16$) & $0$\\\hline
Universal & $0$\\
\botrule
\end{tabular}}
\label{tab:random}
\end{table}

\end{itemize}

\section{Conclusion}

This paper studied the security of an image cipher
based on a variant of Logistic map. Observing its essential
structure, we found that the previous chosen-plaintext attack can be further improved in terms of reducing the number
of chosen plain-images from eight-seven to two and decreasing the computational complexity a little. Beside this,
an effective known-plaintext attack can break the cipher in the sense that only two known plain-image are needed. In addition, some other security flaws, insensitivity to change of plaintext/secret, weak randomness of used PRNG, were identified and briefly discussed.

\section*{Acknowledgement}

This research was supported by the Distinguished Young Scholar Program, Hunan Provincial Natural Science Foundation of China (No.~2015JJ1013), the Natural Science Foundation of China (No. 61202398).

\bibliographystyle{ws-ijbc}
\bibliography{TLM}
\end{document}